\documentclass[%
 reprint,
superscriptaddress,
preprintnumbers,
 amsmath,amssymb,
 aps,
]{revtex4-2}

\usepackage{graphicx}
\usepackage{dcolumn}
\usepackage{bm}
\usepackage{hyperref}
\usepackage{color}
\usepackage[dvipsnames]{xcolor}
\usepackage[
scale=0.85, marginratio={1:1, 2:3}, ignoreall,
]{geometry}
\usepackage{amsmath}
\usepackage{amssymb}
\usepackage{url}
\usepackage{braket}
\usepackage{slashed}
\usepackage[version=4]{mhchem}

\newcommand{\nocontentsline}[3]{}
\newcommand{\tocless}[2]{\bgroup\let\addcontentsline=\nocontentsline#1{#2}\egroup}

\newcommand{\Lag}{\mathcal{L}}

\newcommand{\mpi}{m_{\pi}}
\newcommand{\fpi}{f_{\pi}}

\newcommand{\mnras}{Monthly Notices of the Royal Astronomical Society}
\newcommand{\Rmax}{R_{\text{\tiny max}}^{\text{\tiny stable}}}
\newcommand{\Rmin}{R_{\text{\tiny min}}^{\text{\tiny meta}}}
\newcommand{\mpl}[1]{M^{#1}_\text{\tiny P}}
\newcommand{\RWD}{R_{\text{\tiny WD}}}
\newcommand{\MWD}{M_{\text{\tiny WD}}}

\begin{document}

\DeclareRobustCommand{\Sec}[1]{Sec.~\ref{#1}}
\DeclareRobustCommand{\Secs}[2]{Secs.~\ref{#1} and \ref{#2}}
\DeclareRobustCommand{\App}[1]{App.~\ref{#1}}
\DeclareRobustCommand{\Tab}[1]{Table~\ref{#1}}
\DeclareRobustCommand{\Tabs}[2]{Tables~\ref{#1} and \ref{#2}}
\DeclareRobustCommand{\Fig}[1]{Fig.~\ref{#1}}
\DeclareRobustCommand{\Figs}[2]{Figs.~\ref{#1} and \ref{#2}}
\DeclareRobustCommand{\Figss}[2]{Figs.~\ref{#1} to \ref{#2}}
\DeclareRobustCommand{\Eq}[1]{Eq.~(\ref{#1})}
\DeclareRobustCommand{\Eqs}[2]{Eqs.~(\ref{#1}), (\ref{#2})}
\DeclareRobustCommand{\EqsRange}[2]{Eqs.~(\ref{#1}) - (\ref{#2})}
\DeclareRobustCommand{\Ref}[1]{Ref.~\cite{#1}}
\DeclareRobustCommand{\Refs}[1]{Refs.~\cite{#1}}

\preprint{IFT-UAM/CSIC-22-136}

\title{White dwarfs as a probe of exceptionally light QCD axions}

\author{Reuven Balkin}\email{reuven.b@campus.technion.ac.il}
\affiliation{%
 Physics Department, Technion -- Israel Institute of Technology,  Haifa 3200003, Israel
}%

\author{Javi Serra}\email{javi.serra@ift.csic.es}
\affiliation{%
 Physik-Department, Technische Universit\"at M\"unchen,  85748 Garching, Germany
}%

\affiliation{%
 Instituto de F\'isica Te\'orica UAM/CSIC, Madrid 28049, Spain
}

\author{Konstantin Springmann}\email{konstantin.springmann@tum.de}
\affiliation{%
 Physik-Department, Technische Universit\"at M\"unchen,  85748 Garching, Germany
}%

\author{Stefan Stelzl}\email{stefan.stelzl@epfl.ch}
\affiliation{%
 Physik-Department, Technische Universit\"at M\"unchen,  85748 Garching, Germany
}%

\affiliation{%
 Institute of Physics, Theoretical Particle Physics Laboratory, \\
École Polytechnique Fédérale de Lausanne, CH-1015 Lausanne, Switzerland
}

\author{Andreas Weiler}\email{andreas.weiler@tum.de}
\affiliation{%
 Physik-Department, Technische Universit\"at M\"unchen,  85748 Garching, Germany
}%

\date{\today}

\begin{abstract}
We study the effects of exceptionally light QCD axions on the stellar configuration of white dwarfs. At finite baryon density, the non-derivative coupling of the axion to nucleons displaces the axion from its in-vacuum minimum which implies a reduction of the nucleon mass. This dramatically alters the composition of stellar remnants. In particular, the modifications of the mass-radius relationship of white dwarfs allow us to probe large regions of unexplored parameter space without requiring that axions are dark matter.
\end{abstract}

\maketitle


\section{\label{sec:intro}Introduction}

\begin{figure}[t]
\includegraphics[scale=0.5977]{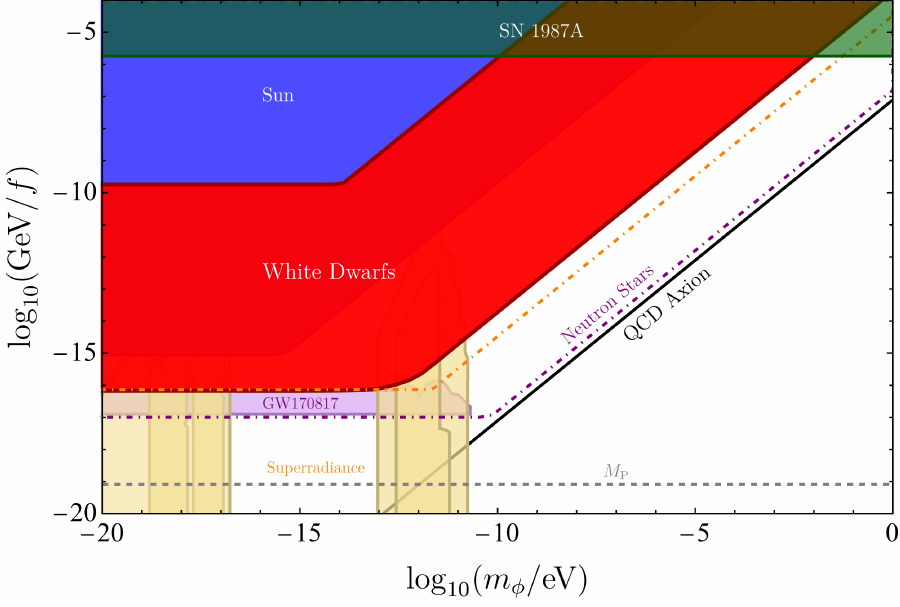}
\caption{\label{fig:Moneyplot} Constraints and future projections on the axion parameter space for the $\epsilon$-model.
Exclusions from modifications of the white dwarf $M\,$-$\,R$ relation are shown in red. Note that the WD bound overlaps with bounds from the Sun in large parts of the region (slightly darker red). The observation of WDs close to the Chandrasekhar limit can further probe the parameter space until the orange dashed line. The solid black line shows the QCD axion with $m_{\phi}f=\mpi\fpi$. For reference, we plot $f = M_{\text{P}}$ in gray. Further bounds originate from the sourcing in the Sun \cite{Hook:2017psm} (blue) and the gravitational wave signal of the NS binary GW170817 \cite{Zhang:2021mks} (violet), which we both adapted at large $f$ according to numerically inspired $\mathcal{O}(1)$ factors, the supernova 1987A \cite{Lucente:2022vuo} (green) , and black hole superradiance \cite{Arvanitaki:2014wva} (yellow). We would furthermore like to note that the pulsar bound of \cite{Hook:2017psm} goes away once all finite gradient effects are properly taken into account (besides lying within a region that is strongly dependent on the neutron star EOS) \cite{Balkin:2023xtr}. Finally, we show which parameters lead to a new ground state accessible in neutron stars (dot-dashed purple); for more details see \cite{Balkin:2023xtr}.}
\end{figure}

Recent years have seen a resurgence of interest in the physics of the QCD axion, driven by a thriving experimental program in sync with a burst of novel theoretical ideas. One instance is the possibility of relaxing the standard relation between the axion potential and its defining couplings to gluons.
Arguably the most exciting outcome of these models is the new set of signals they give rise to beyond the canonical QCD-axion phenomena. 
Here we present a novel implication associated with exceptionally light QCD axions: white dwarfs (WDs) of a certain size should not exist. 
This leads to novel bounds in the QCD axion parameter space, \Fig{fig:Moneyplot}.

We consider models of the QCD axion, a pseudo-scalar field $\phi$ with a coupling to gluons
\begin{equation}
    \Lag=\frac{g_s^2}{32\pi^2}\frac{\phi}{f}G_{\mu\nu}\tilde{G}^{\mu\nu},
    \label{eq:gluon_coupling}
\end{equation}
where $g_s$ is the strong coupling, $f$ is the axion decay constant and $G_{\mu\nu}$ is the gluon field strength. Below the QCD scale the axion obtains the potential~\footnote{In Eqs.~(\ref{eq:potential}) and (\ref{eq:Coupling}) we show approximated expressions using $\cos(\phi/f)$. In practice, we used the more accurate expressions~\cite{GrillidiCortona:2015jxo}
\begin{equation*}
   V = -\epsilon\, m_\pi^2 f_\pi^2 \left(\sqrt{1-\frac{4 m_u m_d}{(m_u+m_d)^2}\sin^2 \left( \frac{\phi}{2f}\right)}-1\right)\,, 
\end{equation*}
and
\begin{equation*}
   \mathcal{L} = -\sigma_N \bar{N} N \left(\sqrt{1-\frac{4 m_u m_d}{(m_u+m_d)^2}\sin^2 \left( \frac{\phi}{2f}\right)}-1\right) \,.
\end{equation*}
}
\begin{equation}
\label{eq:potential}
    V\simeq-\epsilon\,\mpi^2\fpi^2\left[\cos\left(\frac{\phi}{f}\right)-1\right],
\end{equation}
where $\mpi\simeq 135\,\text{MeV}$ is the pion mass and $\fpi\simeq 93\,\text{MeV}$ is the pion decay constant. 
A parameter $\epsilon\leq 1$ is introduced to tune the axion lighter than naively expected~\cite{Hook:2017psm}. 
For a symmetry based realization, see e.g.~\cite{Hook:2018jle,DiLuzio:2021pxd,DiLuzio:2021gos,Banerjee:2022wzk}, where note that around the origin and at finite density, these models are well parameterized by Eq.~(\ref{eq:potential}).
While in the main text we focus on the simple potential \Eq{eq:potential}, a full analysis and discussion of the modifications due to the form of the ZN axion potential \cite{DiLuzio:2021pxd} is presented in \App{app:ZN}.
In vacuum, the axion is stabilized at the origin, thereby solving the strong CP problem.

The coupling to gluons
Eq.~(\ref{eq:gluon_coupling}) also induces an isospin-symmetric coupling to nucleons
\begin{equation}
    \label{eq:Coupling}
    \Lag\simeq-\sigma_N\bar{N}N\left[\cos\left(\frac{\phi}{f}\right)-1\right],
\end{equation}
where $\sigma_N\simeq50\,\text{MeV}$ is the pion-nucleon sigma term. Together with the observed mass, $m_N\simeq939\,\text{MeV}$, this coupling can be interpreted as an effective $\phi$ dependent nucleon mass, $m_N^{\ast}(\phi)\leq m_N$.

A finite nucleon number density,  $\rho_N\equiv\langle \bar{N}\gamma^0 N \rangle$, implies a non-zero expectation value of $\langle \bar{N}N\rangle$. In the non-relativistic limit, these two quantities are approximately equal, $\rho_N\simeq\langle \bar{N}N\rangle$. 

Interestingly, a finite $\rho_N$ can destabilize the axion from its in-vacuum minimum as soon as $\sigma_N\langle\bar{N}N\rangle>\epsilon\,\mpi^2\fpi^2$, with new minima appearing at $\pm \pi f$~\cite{Hook:2017psm,Balkin:2020dsr}. Furthermore, once the axion sits in its new minimum, the neutron mass is reduced by 
\begin{align}
 \delta m_N
    \simeq 32 \,\text{MeV}\left(\frac{\sigma_N}{50\,\text{MeV}}\right).
\end{align}
Reducing the constituent mass acts as additional binding energy in compact objects. 

The sourcing of $\phi$ happens in non-relativistic systems of size $R$ when
\begin{equation}
    R\gtrsim\lambda_{\phi}\simeq 10^{4}\,\text{km}\left(\frac{f}{10^{16}\,\text{GeV}}\right)\left(\frac{10^{-4}\text{MeV}^3}{\rho_N}\right)^{1/2},
\end{equation}
where $\lambda_{\phi}\sim m_{\phi}^{-1}(\rho_N)$ is the typical length scale of the axion at finite density~\cite{Hook:2017psm,Balkin:2021zfd,Balkin:2023xtr}, see also Eq.~(\ref{eq:scale}). Typical WD 
densities fall in the range
$\rho_{\text{WD}}\simeq \left(10^{-4} - 1~\text{MeV}\right)^3$. 
This implies that for systems with a small characteristic length scale, such as single nuclei, 
the axion remains stabilized at $\phi=0$ and therefore does not significantly affect 
forces between nucleons.

However, for large and dense systems such as stellar remnants, the sourcing of $\phi$ implies dramatic changes in their composition and hence their mass-radius ($M\,$-$\,R$) relation. A particularly clean and well-studied example of stellar remnants are WDs. 
The modifications of their $M\,$-$\,R$ relation allow us to probe large regions in the light axion parameter space.
For an exploratory study of axions and other particles influencing the stellar structure of neutron stars in the same manner, see \cite{Gao:2021fyk,Balkin:2023xtr}.

\section{White Dwarf Mass-Radius Relation}

WDs balance the force of gravity with the degeneracy pressure of electrons, while almost their entire mass comes from light but non-relativistic nuclei. Due to charge neutrality, the number density of electrons is related to the energy density of nucleons, $\varepsilon_{\psi}\simeq Y_e m_N\rho_e$, where $Y_e=A/Z$ is the ratio of nucleons per electron. Since WDs 
are composed of light nuclei, ranging from helium \ce{{}^4He} to magnesium \ce{{}^24Mg}, the ratio of nucleons per electron is well-approximated by $Y_e\simeq 2$, see also Fig.~\ref{fig:MR}.
\subsection{Non-interacting gas of electrons and nuclei }
The equation of state (EOS) for a degenerate WD can be described by a Fermi gas of non-interacting electrons together with a gas of nuclei. For simplicity, we take positively charged non-relativistic nuclei, which we denote by $\psi$, with twice the nucleon mass $m_{\psi}=2m_N$.
The pressure is dominated by the electron contribution, $p=p_e+p_{\psi}\simeq p_e$, while the nuclei constitute, to a good approximation, the entire energy density, $\varepsilon=\varepsilon_e+\varepsilon_{\psi}\simeq \varepsilon_{\psi}$. Due to charge neutrality, $\rho_{\psi}=\rho_e\equiv\rho$, 
we relate the electron Fermi momentum to 
the energy density as $k_F=(3\pi^2\varepsilon_{\psi}/Y_e m_N)^{1/3}$, and
derive the EOS
\begin{equation}
    \label{eq:WDEOS}
    p(\varepsilon)=\frac{2}{3}\int_0^{k_F(\varepsilon)}\frac{\text{d}^3k}{\left(2\pi\right)^3}\frac{k^2}{\sqrt{k^2+m_e^2}}.
\end{equation}

\begin{figure}[t]
\includegraphics[scale=0.5977]{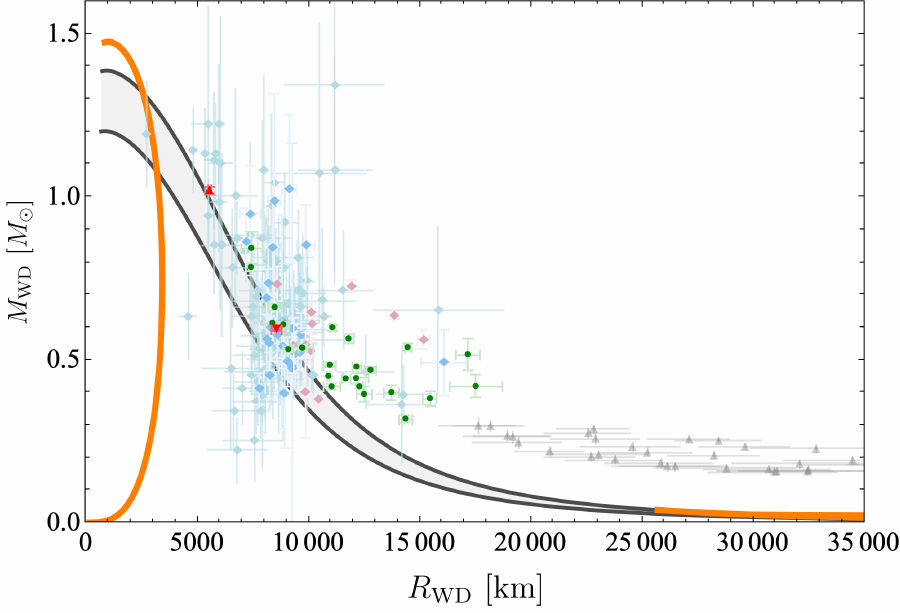}
\caption{\label{fig:MR} White dwarf $M\,$-$\,R$ relation with light QCD axions.  
Free Fermi gas of nuclei and electrons without an axion (black). 
The upper and lower bands correspond to the constitutions of light and more heavy nuclei, i.e.~\ce{{}^4He} which corresponds to $Y_e=2$ and \ce{{}^56Fe} corresponding to $Y_e=2.15$ respectively, while the gray shaded area corresponds to intermediate values.
In orange we show the two branches with an axion for $\epsilon=10^{-11}$ in the limit $\RWD\gg \lambda_{\phi}$ for $Y_e=2$.
The meta-stable branch follows the free Fermi gas line at large radii, while the new ground state phase has much smaller radii. 
Data points are taken from \cite{B_dard_2017} (turquoise), \cite{tremblay_gaia_2016} (blue), \cite{2018MNRAS.479.1612J} (pink), \cite{Bond_2015,Bond_2017} (red), \cite{10.1093/mnras/stx1522} (green) and \cite{Brown_2020} (gray). One can clearly see the gap in the predicted $M\,$-$\,R$ relation that is incompatible with data.}
\end{figure}

We work in the limit $T\rightarrow 0$, which is justified since $T/\mu_e\ll 1$ for typical WDs, where $\mu_e$ is the electron chemical potential.
Temperature shifts the $M\,$-$\,R$ relation
up to higher masses, an effect most relevant for the largest and most dilute WDs, see Refs.~\cite{10.1093/mnras/stx1522,Nunes:2021jdt,PhysRevC.89.015801,1972A&A....16...72S,1970A&A.....8..398K,cite-key,1975ApJ...200..306L} as well as Ref.~\cite{1990RPPh...53..837K} for a review.

The EOS completes the set of equations that describe the balance between the electron degeneracy pressure and gravity, the Tolman-Oppenheimer-Volkoff (TOV) equations~\cite{Oppenheimer:1939ne,Tolman:1939jz}, 
\begin{subequations}
\label{eq:TOV}
\begin{align}
&p'= -\frac{GM\varepsilon}{r^2}\bigg[1+\frac{p}{\varepsilon}\bigg]\left[1-\frac{2GM}{r}\right]^{-1}\left[1+\frac{4\pi r^3 p}{M}\right], \\
&M'=   \, 4\pi r^2 \varepsilon,
\end{align}
\end{subequations}
where $G=M_{\text{P}}^{-2}$ is Newton's constant, $M(r)$ the enclosed mass and all derivatives are taken with respect to the radial coordinate.
Solutions with varying central
pressures lead to a $M\,$-$\,R$ relationship that agrees well with current data, as shown in Fig.~\ref{fig:MR}.

\subsection{The axion WD system: a new ground state}
In the presence of the axion, the full system is described by a free Fermi gas of electrons, an ideal gas of nuclei with a $\phi$-dependent mass~\footnote{The interesting case of a scalar coupled non-derivatively to electrons inducing a $\phi$ dependent mass $m^{\ast}_e(\phi)$ is left for a separate publication \cite{Balkin:20222} }, $m_{\psi}^{\ast}(\phi)=2m^{\ast}_N(\phi)$, 
the gravitational field $g_{\mu\nu}$, and the axion $\phi$.
The gravitational field is sourced by an energy-momentum tensor 
\begin{equation}
    T_{\mu\nu}=T^{\psi\phi}_{\mu\nu}+T^{\text{grad}}_{\mu\nu}.
    \label{eq:energystress}
\end{equation}
The first term takes the form of an ideal fluid,
 \mbox{$T^{\psi\phi}_{\mu\nu}=\text{Diag}\left(\varepsilon,-p,-p,-p\right)$}
, with
\begin{align}
    p(\phi,\rho)&=\frac{2}{3}\int_0^{k_F(\rho)}\frac{\text{d}^3k}{\left(2\pi\right)^3}\frac{k^2}{\sqrt{k^2+m_e^2}}-V(\phi),
    \\
    \varepsilon(\phi,\rho)&=m_{\psi}^{\ast}(\phi)\rho+\varepsilon_e(\rho)+V(\phi),
\end{align}
where we neglected the sub-leading contributions to the pressure $p_{\psi}(\phi,\rho) \ll p_{e}(\rho)$.
The second term in Eq. (\ref{eq:energystress}) contains the contribution of the axion gradient
\begin{equation}
    {(T^{\text{grad}})^{\mu}}_{\nu}=\frac{(\phi')^2}{2}\left[1-\frac{2GM}{r}\right]\left(\delta^{\mu}_{\nu}-2\delta^{\mu}_{r}\delta^{r}_{\nu}\right).
\end{equation}
Using Einstein's and the scalar equations of motion, we find the following set of coupled differential equations
\begin{widetext}
\begin{subequations}
    \label{eq:coupledTOV}
\begin{align}
    &\phi''\bigg[1-\frac{2GM}{r}\bigg]+\frac{2}{r}\phi'\left[1-\frac{GM}{r}-2\pi Gr^2\left(\varepsilon-p\right)\right]
    =\frac{\partial V}{\partial \phi}+\rho\frac{\partial m_{\psi}^{\ast}(\phi)}{\partial \phi}\equiv U(\phi,\rho), \label{eq:AxionEOM} \\
    &p'=-\frac{GM\varepsilon}{r^2}\bigg[1+\frac{p}{\varepsilon}\bigg]\left[1-\frac{2GM}{r}\right]^{-1}\left[1+\frac{4\pi r^3}{M}\left(p+\frac{(\phi')^2}{2}\left\{1-\frac{2GM}{r}\right\}\right)\right]
    -\phi'U(\phi,\rho),\label{eq:coupledTOV1}\\
    &M'=\,\,4\pi r^2\left[\varepsilon+\frac{1}{2}\left(1-\frac{2GM}{r}\right)\left(\phi'\right)^2\right]\label{eq:coupledTOV2}.
\end{align}
\end{subequations}
\end{widetext}
Eq.~(\ref{eq:AxionEOM}) is the static axion equation of motion coupled to gravity, while Eqs.~(\ref{eq:coupledTOV1}) and 
(\ref{eq:coupledTOV2}) are the TOV equations in the presence of an axion. 
Note that, we recover the ordinary TOV equations, Eq.~(\ref{eq:TOV}), in the limit $\phi=0$. While it is possible to numerically solve Eq.~(\ref{eq:coupledTOV}) using the shooting method, there exists a limit in which these equations simplify dramatically. 

The displacement of the axion at sufficiently high densities costs gradient energy and therefore it only occurs if balanced by the gain in potential energy. This leads to the typical scale on which the axion is displaced
\begin{equation}
    \lambda_{\phi}(\rho)\simeq\frac{\pi f}{\sqrt{2(\delta m_N \rho- \epsilon m_{\pi}^2 f_{\pi}^2})},\label{eq:scale}
\end{equation}
to be evaluated at typical WD densities.

For $\RWD \gg  \lambda_{\phi}$, the field essentially tracks the minimum of the effective in-density potential on stellar scales and is given by the solution to
\begin{equation}
    U(\phi,\rho)=0. \label{eq:minimum}
\end{equation}
At the same time, the gradient terms in Eqs.~(\ref{eq:coupledTOV1}) and 
(\ref{eq:coupledTOV2}) are confined to a small transition shell, where the field does not follow its minimum. However, this localized contribution is negligible as long as $\frac{\lambda_{\phi}}{\RWD} \frac{\delta m_N}{m_N} \ll 1$, which is trivially fulfilled in this case.

Therefore, for large systems we can neglect the axion gradient $\phi'\simeq0$. As a result, Eq.~(\ref{eq:coupledTOV})  decouples to give the regular TOV equations, Eq.~(\ref{eq:TOV}), in addition to Eq.~(\ref{eq:minimum}).
Note that the latter is the same condition as the minimization of the energy density $\varepsilon(\phi,\rho)$ with respect to $\phi$. Solutions $\phi(\rho)$ describe a thermodynamically stable EOS used to solve the regular TOV equations.

Interestingly, if the axion is destabilized in a WD, the energy per particle of the light nuclei
$\varepsilon(\rho)/\rho$
is not minimized when the nuclei are infinitely separated ($\rho \to 0$), but rather at some finite density $\rho^{\ast}$, which can be found numerically. This implies the existence of an energetically favored state of matter at $\rho^{\ast}$, where the axion is at $\braket{\phi}=\pm\pi f$. This new ground state is in fact reminiscent to strange quark matter~\cite{Witten:1984rs}. Note that the density of the new ground state is slightly larger than the density at which the destabilization occurs, $\rho^{\ast}>\rho_c \equiv \epsilon\,\mpi^2\fpi^2/2\sigma_N$. 

\begin{figure}[t]
\includegraphics[scale=0.33]{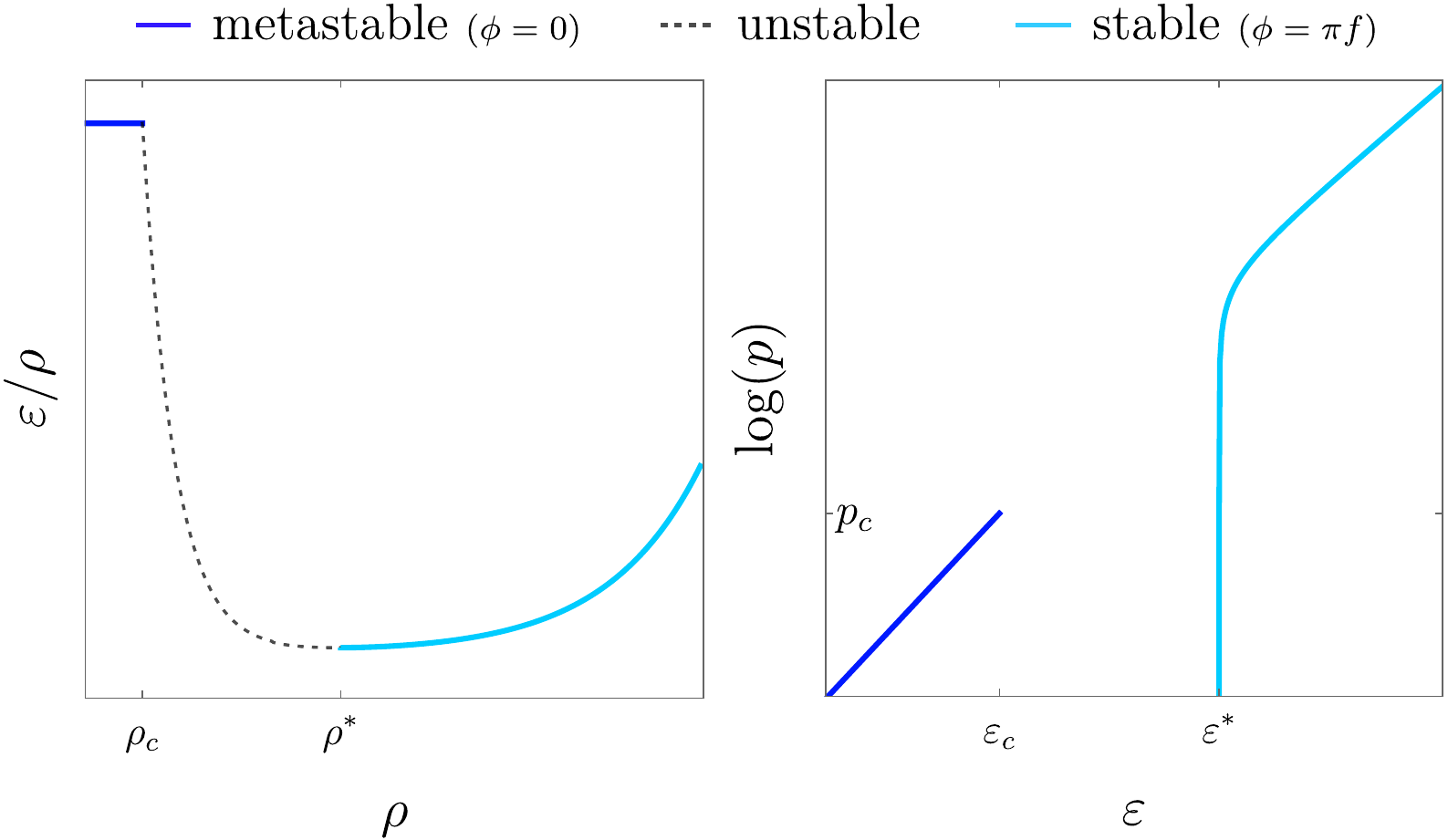}
\caption{\label{fig:EOSplot} The energy per particle $\varepsilon/\rho$ as a function of number density (left) and the EOS (right). At low densities $\rho<\rho_c$, the system is in its meta-stable $\phi=0$ phase (dark blue). For $\rho_c<\rho<\rho^*$, the system is unstable i.e.~$p<0$ (dashed line in left panel). At larger densities 
$\rho^*<\rho$, the system is in its $\phi=\pi f$ phase (light blue), with a new ground state at $\varepsilon^*$ where $p=0$. }
\end{figure}

For low densities $\rho<\rho_c$, matter is in a meta-stable state where the classical sourcing of the axion is not preferable.
Once $\braket{\phi}=\pm\pi f$, there is a range $\rho_c <\rho<\rho^{\ast}$, where the energy per particle decreases $\partial_{\rho}(\varepsilon(\rho)/\rho) <0$, implying a negative pressure. At densities slightly above $\rho_c$,
the total pressure turns negative due to the onset of the the axion potential $p=p_e-V<0$. As $\rho$ increases, $V$ stays constant while $p_e$ increases, until finally the system stabilizes at
$p=p_e(\rho^{\ast})-V=0$, see Fig.~\ref{fig:EOSplot}. In this unstable phase the system contracts until it stabilizes in the new ground state. 

This instability leads to a \emph{gap} in the predicted $M\,$-$\,R$ relationship as seen in Fig.~\ref{fig:MR}. The position of this gap is $\epsilon$ dependent; the smaller $\epsilon$ is, the more the gap is shifted towards small masses and large radii. We use the position of this gap to probe the existence of light QCD axions.

Note that the simplified discussion above is only valid for $R_\text{WD}\gg\lambda_{\phi}$. For $R_\text{WD}\sim\lambda_{\phi}$ we numerically solve the full coupled system in Eq.~(\ref{eq:coupledTOV}) and find that for large values of the axion decay constant, $f$
, and small $\epsilon$, the position of the gap is $\epsilon$ independent.
This is understood as follows: on the stable branch, the gradient pressure, which is controlled by $f$, 
is relevant. If gravity is subdominant, this pressure fixes the central density of the star,  \mbox{$\rho(r=0)>\rho^{\ast}$}.
The maximal radius is then achieved when the gravitational pressure equals the gradient pressure. For the meta-stable branch, the minimum radius is set by $R \sim \lim_{\epsilon\to 0}\lambda_\phi$.

Finally, in the limit $R_\text{WD}\ll\lambda_{\phi}$, the gradient energy is so large that the field cannot move away from its in-vacuum minimum and therefore has no influence on the structure of WDs.

\subsection{Confrontation with Observational Data}
There are large data sets available containing masses and radii of WDs (see e.g.~\cite{Gentile_Fusillo_2021,refId1,refId0,Koester2019,2018MNRAS.480.4505J,tremblay_gaia_2016,2018MNRAS.479.1612J,B_dard_2017,Bond_2015,Bond_2017,10.1093/mnras/stx1522,Bergeron_2019,Genest_Beaulieu_2019,Brown_2020}). However, not all of these data-sets can be used to probe the $M\,$-$\,R$ relation. In some catalogs, (see \cite{Gentile_Fusillo_2021,refId1,refId0,Koester2019,2018MNRAS.480.4505J}), the $M\,$-$\,R$ relation is used as an input to significantly reduce observational error. On the other hand, there are sets (e.g.~\cite{10.1093/mnras/stx1522,tremblay_gaia_2016,2018MNRAS.479.1612J,B_dard_2017,Bond_2015,Bond_2017,Bergeron_2019,Genest_Beaulieu_2019,Brown_2020}) that systematically test the $M\,$-$\,R$ relation using observational data. While in \cite{10.1093/mnras/stx1522} the determination of the mass and radius is completely independent of WD models, most other works depend on an atmospheric model to determine the radius. Nevertheless, we combine the data sets \cite{10.1093/mnras/stx1522,tremblay_gaia_2016,2018MNRAS.479.1612J,B_dard_2017,Bond_2015,Bond_2017,Brown_2020} and show them in Fig.~\ref{fig:MR}. 

The data of measured WD masses and radii is scattered broadly between radii of $(5000 - 40000 ~\text{km})$, which matches reasonably well with the free Fermi gas description. The notable deviation in mass found at large radii in Fig.~\ref{fig:MR} is due to finite temperature effects; $\mu_e$ in these dilute stars is typically smaller, increasing the relevance of $T/\mu_e$ corrections.
Finite temperature effects lead to modifications of the EOS and to a slight modification of masses and radii, but still predict a continuous $M\,$-$\,R$ curve. 
The same holds for other well-known corrections to the EOS, such as different compositions, electrostatic corrections, or nuclear reactions, see e.g.~Refs.~\cite{Shapiro:1983du,1990RPPh...53..837K}. While nuclear reactions change in the sourced phase e.g.~due to a different mass of the pions, this has negligible effect on the static structure of white dwarfs \cite{Ubaldi:2008nf} \footnote{In this work we take a conservative approach and stay agnostic of the process of formation of the new ground state phase, where the change of nuclear reactions could play a role.}.

We perform a simplified statistical analysis to determine the compatibility of the observed WD radii with a gapped radius distribution hypothesis (marginalizing over mass and neglecting small theory systematics). 
We summarize here its main results, with the full details given in the \App{sup::stat_and_bounds}.  
For the purpose of the analysis, we calculate the position of the radius gap as a function of $\epsilon$ and $f$, relying both on numerical results, as well as on numerically-verified analytical estimates. 
In the region $f \ll 10^{9}\,$GeV, finite gradient effects are negligible w.r.t.~the position of the gap, making it $f$-independent. For the simplified model described in the main text, we are able to exclude at the $2\sigma$ level the following interval in $\epsilon$,
\begin{align}
2\times10^{-20}  \lesssim \epsilon \lesssim 2\times 10^{-7} \;\;\;\; \text{(95\% CL)}\,,
\end{align}
see \Fig{fig:Moneyplot}. On the other hand, for the $Z_\mathcal{N}$-model, our analysis (see \App{app:ZN}) leads to the exclusion of
\begin{align}
33 \le \mathcal{N} \le 69 \;\;\;\; \text{(95\% CL)}\,,
\end{align}
as shown in \Fig{fig:MoneyplotZN}.

The upper limit is set by the smallest, most massive WDs and our analysis effectively excludes all points in the axion parameter space that cannot predict a WD with a radius smaller than around $\sim 4000\,\text{km}$ on the meta-stable branch.
The lower limit is sensitive to the largest, extremely low-mass WDs \cite{Brown_2020}.
For even lower values of $\epsilon$, the stable branch covers most of the range of observed radii. 
In this case, although all the observed WDs are sourcing the scalar field, the gap in radius is relegated to extremely large (and potentially unpopulated) WD radii. 
Note however that this region in parameter space is already ruled out by requiring that no sourcing occurs in our sun~\cite{Hook:2017psm}.

Conversely, in the region $\epsilon \ll 10^{-20}$, or $\mathcal{N}\gg 71$, finite gradient effects are dominant w.r.t.~the position of the gap, making it $\epsilon$-independent.
Using solutions of the coupled system, Eq.~\ref{eq:coupledTOV}, we are able to exclude at the $2\sigma$ level the following interval in $f$,
\begin{align}
5.5\times10^{9} < f/\text{GeV} < 1.1\times 10^{16}\;\;\;\; \text{(95\% CL)}\,,
\end{align}
see Fig.~(\ref{fig:Moneyplot}).
The upper value represents the limit in which WDs are not large enough to source the scalar field, i.e. $\lambda_\phi \gtrsim R_{\text{\tiny WD}}$. 
In the region $f \epsilon^{-1/3} \sim \mpl{}$, in which both gradient and finite $\epsilon$ effects are important, we verify numerically that the sourcing stops at lower values of $f$. 
Similarly to the lower bound on $\epsilon$, the lower bound on $f$ is sensitive to the largest, extremely low-mass WDs.
We stress again that we do not expect our results to strongly depend on finite temperature effects.

\section{Conclusions}
The mass-radius relationship of white dwarfs is well-understood and has been observationally tested with increasing accuracy in recent years. 
We showed how light QCD axions change the structure of WDs, thus predicting the presence of a gap. We used existing data to place novel bounds on their parameter space. We stress that the bounds arising from the existence of a new ground state accessible in white dwarfs, and the corresponding gap in radii, are qualitatively very different than the strategy proposed in Ref. \cite{Hook:2017psm}, which relies on the change of the properties of nuclei, and the corresponding change in X-ray emission, when a (lighter) QCD axion is displaced to $\theta=\pi$ \cite{Ubaldi:2008nf}.

The QCD axion generically predicts a non-derivative coupling to nucleons. At finite baryon density this coupling can destabilize the axion from its in-vacuum minimum. If sourced, the non-zero axion expectation value reduces the mass of nucleons. For a large region of the parameter space, this leads to a new ground state of matter, which has less energy per particle than infinitely separated nucleons. If accessible in WDs, this drastically changes their $M\,$-$\,R$ relation. Since the axion is sourced by the WD, this does not rely on the axion contributing to the dark matter relic abundance.

More precise tests of the WD $M\,$-$\,R$ curve using the recent Gaia DR3 are expected in the near future and will further probe the parameter space of light QCD axions.

As a consequence of the new ground state of matter, we predict new small self-bound objects held together by the gradient pressure of the axion. These objects could give rise to novel signatures of exceptionally light QCD axions down to the QCD axion line.

\begin{acknowledgments}
The work of JS, KS, SS and AW has been partially supported by the Collaborative Research Center SFB1258, the Munich Institute for Astro- and Particle Physics \mbox{(MIAPP)}, and by the Excellence Cluster ORIGINS, which is funded by the Deutsche Forschungsgemeinschaft (DFG, German Research Foundation) under Germany’s Excellence Strategy – EXC-2094-390783311. 
The work of JS is supported by the grant RYC-2020-028992-I funded by MCIN/AEI /10.13039/501100011033 and by ``ESF Investing in your future''. JS also acknowledges the support of the Spanish Agencia Estatal de Investigacion through the grant ``IFT Centro de Excelencia Severo Ochoa CEX2020-001007-S''.
The work of SS is additionally supported by the Swiss National Science Foundation under contract 200020-18867. 
The work of RB is supported by grants from the NSF-BSF (No. 2018683), the ISF (No. 482/20), the BSF (No. 2020300) and by the Azrieli foundation. 
KS would like to thank the elementary particle physics group at the University of Maryland for their hospitality during the final stages of this work.
\end{acknowledgments}

\appendix

\section{Analytic estimates for the radius gap}
\label{sup::stat_and_bounds}
We define $p_0$ as the inward pointing pressure at the core of a white dwarf (WD) as a sum of a gravitational and a gradient contributions
\begin{align}
p_0 &\simeq 
\Delta p_{\text{\tiny grav}}+p_{\text{\tiny grad}}   \,,
\label{eq::master_eq}
\end{align}
with the gravitational pressure is given by
\begin{align}
\Delta p_{\text{\tiny grav}} = \frac{ R^2 m_N^2 \rho^2_0 
}{\mpl{2}}\,, 
\label{eq::pressure_grav}
\end{align}
where $\rho_0$ is the number density at the core ($r=0$). 
The gradient pressure in the sourced phase (i.e. stars on the stable branch) is given by
\begin{align}
p_{\text{\tiny grad}} = \begin{cases}
 \frac{f^2}{R \lambda_{\phi}} =  \frac{ f\sqrt{ \delta{m_N} \rho_R }}{R } \;\;\;\;\;\;  &\lambda_{\phi} \ll R
\\
\frac{f^2}{R^2} & \lambda_{\phi} \sim R
\end{cases}\,,
\label{eq::pressure_grad}
\end{align}
where $\rho_R $ is the number density at the edge ($r=R-\lambda_\phi \approx R$) of the WD. The first line in Eq.~(\ref{eq::pressure_grad})  represents the thin wall limit $\lambda_{\phi} \ll R$, in which the gradient pressure is exerted at a small transition region at the edge of the star. 
In the last step we use the definition of the in-medium wavelength of Eq.~(\ref{eq:scale}) assuming a negligible contribution from the scalar potential and neglecting $\mathcal{O}(1)$ factors.
The second line in Eq.~(\ref{eq::pressure_grad}) is the opposite regime $\lambda_{\phi} \sim R$, where the gradient pressure is delocalized and is spread throughout the star. This is the typical edge case configuration in which the star is barely large enough to source the  axion. In the unsourced phase, i.e.~stars on the meta-stable branch, $p_{\text{\tiny grad}}=0$. See Ref.~\cite{Balkin:2023xtr} for more details on the derivation of Eqs.~(\ref{eq::pressure_grav}) and (\ref{eq::pressure_grad}).

\begin{figure*}[t]
\includegraphics[width=0.49\textwidth]{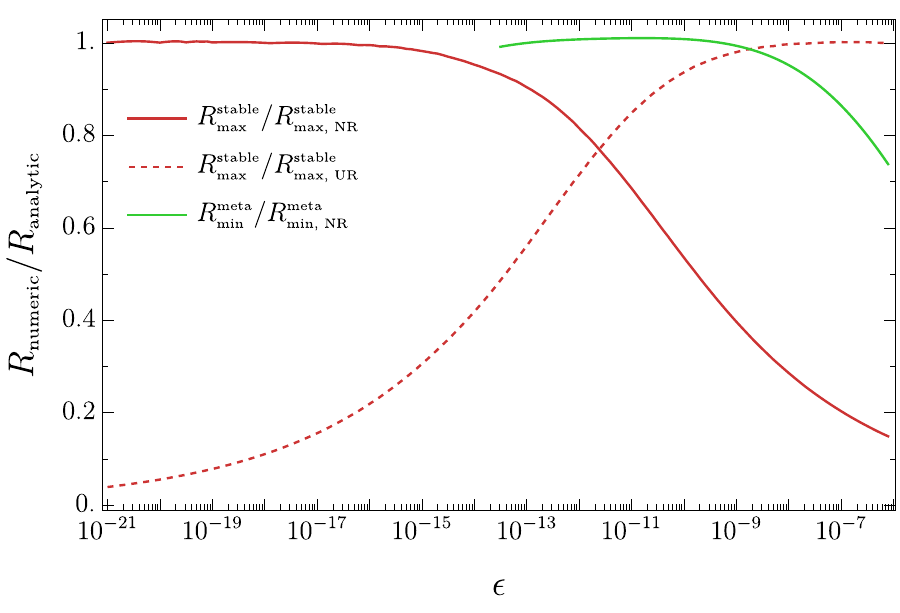}
\includegraphics[width=0.49\textwidth]{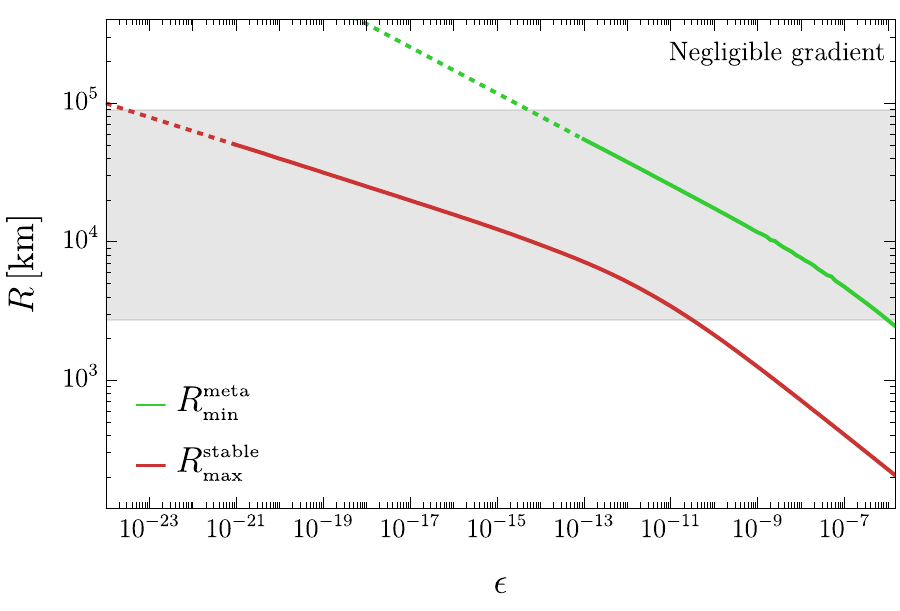}
\caption{Left panel: the ratio between the numerical results and the analytical estimates for the radii which correspond to the edges of the radius gap, as a function of $\epsilon$. In (dashed) red we plot the ratio for  $\Rmax$ divided by the NR (UR) estimate given in Eq.~(\ref{eq::Rmax_f0}). In green we plot the ratio for $\Rmin$, divided by the NR estimate give in Eq.~(\ref{eq::Rmin_f0}). In all cases we match the $\mathcal{O}(1)$ prefactors to the numerical results.  Right panel:   the radius gap defined by $\Rmax$ (red) and $\Rmin$ (green) as a function of $\epsilon$. Solid curves indicate region where numerical results are used while the dashed curve indicates where extrapolation (using the verified analytic estimates) is used. The gray region corresponds to the observed radii range of WDs.}
\label{fig:Rnum_Ranalytic_f0}
\end{figure*}

On the other hand, we define the outwards pressure at the core balancing $p_0$, see Eq. \ref{eq::master_eq}, as the contribution of the electron gas and the scalar potential, which can be written analytically in the non-relativistic (NR) and ultra-relativistic (UR) limits as
\begin{align}
p_0 & \simeq -V(\phi)+p_e(\rho_0)\,, \\
p_e(\rho_0) & \simeq  \begin{cases} \rho_0^{5/3}/m_e\;\;\;\;\;\;\;\;& \text{(NR $\rho_0 \ll m_e^3$)}
\\
\rho_0^{4/3} & \text{(UR $\rho_0 \gg m_e^3$)}
\label{eq::electron_pressure}
\end{cases}\,,
\end{align}
where in the sourced phase we have by definition $V(\phi) = p_e(\rho^*)$, while in the unsourced phase the scalar potential vanishes $V(\phi)=0$.
Note that in Eqs.~(\ref{eq::pressure_grav}), (\ref{eq::pressure_grad}) and (\ref{eq::electron_pressure}), we work at leading order in $\delta{m_N}/m_N \ll 1$ and neglect $\mathcal{O}(1)$ numerical prefactors.

Let us start by estimating the minimal radius on the meta-stable branch, which we denote by $\Rmin$, and the maximal radius on the stable branch, which we denote by $\Rmax$, in the negligible gradient regime, where $\Delta p_{\text{\tiny grav}} \gg p_{\text{\tiny grad}}$. 
In this limit, $\Rmax$ is the radius of the largest approximately constant energy density configurations. Therefore, we estimate it by setting $\rho_0 \approx c\,\rho^* \approx c\,\epsilon^{3/5}(m_e \mpi^2 \fpi^2)^{3/5}$, where $c \sim \mathcal{O}(1)$, and solving for $R$. 
The contribution to the pressure from $V(\phi) = p_e(\rho^*)$ is neglected since it is at most of the same order as $p_e(\rho_0)$, and would therefore have at most an $\mathcal{O}(1)$ effect on the final result. 
We find
\begin{align}
\Rmax(\epsilon) = \frac{\mpl{}}{m_N \Lambda_{\text{\tiny QCD}}} \begin{cases}
\left(\frac{\Lambda_{\text{\tiny QCD}}}{m_e}\right)^{3/5}\epsilon^{-1/10} & \text{(NR)}
\\
\epsilon^{-1/4} & \text{(UR)}
\end{cases}\,,
\label{eq::Rmax_f0}
\end{align}
where for brevity we denoted $\mpi^2 \fpi^2 \equiv \Lambda^4_{\text{\tiny QCD}}$. 
We omit the weak dependence on $c$, which amounts to an $\mathcal{O}(1)$ pre-factor.
In the left panel of Fig.~(\ref{fig:Rnum_Ranalytic_f0}) we compare the analytic estimates to the numeric results. 
We find that the NR estimation is in excellent agreement with the numerical results for $\epsilon \lesssim 10^{-13} $ (red curve). 
For larger values of $\epsilon$, the smaller minimal radii on the stable branch correspond to denser configurations, where relativistic corrections become important. 
Thus, for $\epsilon \gtrsim 10^{-11} $ the UR estimation agrees with the numerical results (red dashed curve).

The edge of the meta-stable branch $\Rmin$ is found by taking $\rho_0 \approx \rho_{c} \approx \epsilon\Lambda_{\text{\tiny QCD}}^4/(2\sigma_N)$ and solving for $R$ in the NR approximation, resulting in
\begin{align}
\Rmin(\epsilon) = \frac{\mpl{}}{m_N \Lambda_{\text{\tiny QCD}}}\left(\frac{\sigma_N \Lambda_{\text{\tiny QCD}} ^2 }{  m_e^3 }\right)^{1/6}\epsilon^{-1/6} \;\;\;\;\; (\text{NR})\,.
\label{eq::Rmin_f0}
\end{align}

A similar UR approximation is straightforward to derive. However, it is only valid for $R \ll \mpl{}/(m_N m_e) \sim 5000\,$km, which is outside our range of interest. In the left panel of Fig.~(\ref{fig:Rnum_Ranalytic_f0}), we compare the analytic numerical results of $\Rmin(\epsilon)$ to the analytic estimate. We find good agreement in most of the calculated region, namely for $\epsilon \lesssim 10^{-9}$. For larger values, relativistic corrections start becoming important and the UR approximation begins to degrade. 

The radius gap in negligible gradient regime is plotted in the right panel of Fig.~(\ref{fig:Rnum_Ranalytic_f0}) as a function of $\epsilon$. For the purpose of the analysis of Sec.~(\ref{sup::stat_and_bounds}), solid curves indicate the regions where we use our numerical results, while dashed lines indicates regions where extrapolation, based on the verified numerical estimate, is used.  The gray region corresponds to the observed radii range of WDs.\\

In the region of parameter space where $\epsilon$ is negligibly small (to be determined below),
the position of the radius gap is determined by finite gradient effects.  
On one side, the edge of the meta-stable branch indicates when a region of size $\lambda_\phi$ with above-critical density is formed, which leads to an instability. 
On the other side, the largest configurations on the stable branch are those in which the gravitational pressure begins to dominate over the gradient pressure exerted at the edge of the star~\cite{Balkin:2023xtr}.

\begin{figure*}[t]
\includegraphics[width=0.49\textwidth]{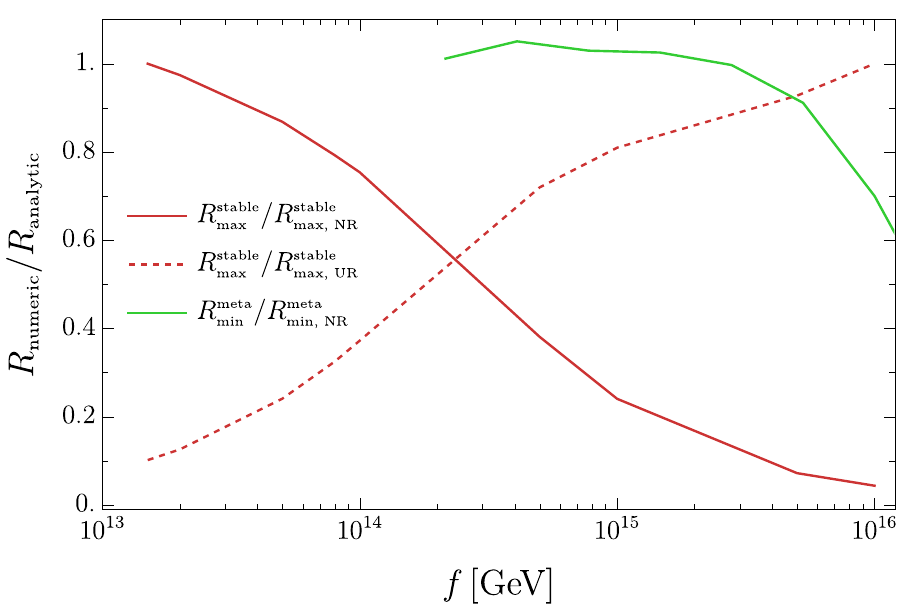}
\includegraphics[width=0.49\textwidth]{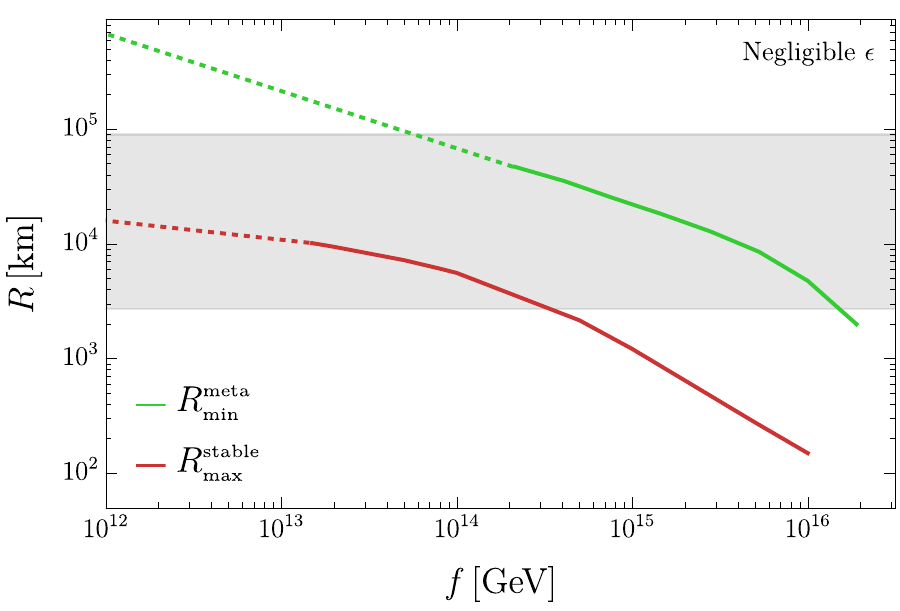}
\caption{Left panel: the ratio between the numerical results and the analytical estimates for the radii which correspond to the edges of the radius gap, as a function of $f$. In red and dashed-red we plot the ratio for  $\Rmax$ divided by the NR and UR estimates given in Eq.~(\ref{eq::Rmax_NR_eps0}) and Eq.~(\ref{eq::Rmax_UR_eps0}), respectively. In green we plot the ratio for $\Rmin$, divided by the NR estimate give in Eq.~(\ref{eq::Rmin_eps0}). In all cases we match the $\mathcal{O}(1)$ prefactors to the numerical results.  Right panel:   the radius gap defined by $\Rmax$ (red) and $\Rmin$ (green) as a function of $f$. Solid curves indicate region where numerical results are used while the dashed curve indicates where extrapolations (using the verified analytic estimates) are used. The gray region corresponds to the observed radii range of WDs.}
\label{fig:Rnum_Ranalytic_eps0}
\end{figure*}

First we find $\Rmax$ for lower values of $f$ where the thin-wall approximation holds. by taking $\rho_0=\rho_R\equiv \rho_{\text{\tiny eq}}>\rho^* $, where $ \rho_{\text{\tiny eq}}$ is found by solving $\Delta p_{\text{\tiny grav}}(\rho_{\text{\tiny eq}} )=\Delta p_{\text{\tiny grav}}(\rho_{\text{\tiny eq}} )$ for $\rho_{\text{\tiny eq}}$. 
We than plug $\rho_{\text{\tiny eq}}$ into Eq.~(\ref{eq::master_eq}), and using the NR approximation of Eq.~(\ref{eq::electron_pressure}) solve for $R$ and find
\begin{align}
\Rmax(f) 
=
\left( \frac{\mpl{}}{m_N}\right)^{7/6}\left(\frac{1}{\delta m_N ^{1/12} f^{1/6} m_e^{3/4} }\right) \;\;\;\; (\lambda_\phi \ll R)\,,
\label{eq::Rmax_NR_eps0}
\end{align}
where we again neglect the contribution from $V(\phi)$ as an $\mathcal{O}(1)$ correction at most.
We compare this estimate with the numerical results in Fig.~(\ref{fig:Rnum_Ranalytic_eps0}) (red curve). 
We find is is consistent with the numerical results in the region $f \ll 10^{15}\,$GeV. 
Above these values of $f$, the thin-wall approximation breaks down and $\Rmax(f)$ can be estimated using the $\lambda_\phi \sim R$ expression for $p_{\text{\tiny grav}}$ and the UR expression for the electron pressure, which gives us
\begin{align}
\Rmax(f) = \frac{\mpl{2}}{m_N^2 f} \;\;\;\; (\lambda_\phi \lesssim R)\,.
\label{eq::Rmax_UR_eps0}
\end{align}
We find this estimate consistent with the numerical results in the region $f \gg 10^{15}\,$GeV, see dashed curve in Fig.~(\ref{fig:Rnum_Ranalytic_eps0}). 

The edge of the meta-stable branch $\Rmin$ is found by first finding the critical density for which the while size of the star is of the order of the scalar in-medium wavelength, namely by solving $\lambda_\phi(\rho) = f/\sqrt{\rho \delta m_N}=R$ for $\rho$ and plugging the result in Eq.~(\ref{eq::master_eq}) using the NR approximation for the electron gas. We find
\begin{align}
\Rmin(f) = \left(\frac{\delta m_N^{1/2} \mpl{3} }{f m_e^{3/2} m_N^{3}}\right)^{1/2}\,.
\label{eq::Rmin_eps0}
\end{align}
We find this estimate consistent with the numerical results, see green curve in Fig.~(\ref{fig:Rnum_Ranalytic_eps0}). The deviations from the analytical estimate at large value of $f$ is expected, since for smaller and denser stars relativistic corrections to the EOS become increasingly larger. 

To conclude we note that for the whole $\{\epsilon,f\}$ plane, we define the maximal radius of the gap as
\begin{align}
\Rmin(\epsilon,f) = \text{Min}\left[\Rmin(\epsilon),\Rmin(f) \right]\,,
\end{align}
where the two radii coincide $\Rmin(\epsilon)\sim\Rmin(f) $ around the curve defined by
\begin{align}
f \epsilon^{-1/3} \approx \frac{ \sqrt{\delta m_N} }{\sqrt{m_e} m_N}\left(\frac{\Lambda^4_{\text{\tiny QCD}}}{\sigma_N}\right)^{1/3}\mpl{} \sim \mpl{} \,,
\end{align}
using the NR estimations for both expressions.
This defines (a posteriori) the ranges of validity for the negligible gradient and negligible $\epsilon$ approximations for the determination of  $\Rmin$, e.g. the negligible gradient is a valid approximation in the region of parameter space where $f \epsilon^{-1/3} \ll \mpl{}$. Around $f \epsilon^{-1/3} \sim \mpl{}$, we computed a numerical solution using the appropriate $\{\epsilon,f\}$ values in order to obtain $\Rmin(\epsilon,f)$. 
In a similar fashion, we define for the whole $\{\epsilon,f\}$ plane the minimal radius of the gap as
\begin{align}
\Rmax(\epsilon,f) = \text{Max}\left[\Rmax(\epsilon),\Rmax(f) \right]\,,
\end{align}
where the two radii coincide $\Rmax(\epsilon)\sim\Rmax(f) $ around the curve defined by
\begin{align}
f \epsilon^{-3/5} \approx \frac{\Lambda_{\text{\tiny QCD}}^{12/5} \mpl{}}{\sqrt{\delta m_N} m_e^{9/10} m_N} \sim 20 \mpl{} \,,
\end{align}
using the NR estimations for both expressions.
This defines (a posteriori) the ranges of validity for the negligible gradient and negligible $\epsilon$ approximations for the determination of  $\Rmax$.

\section{Statistical analysis and bounds}
\label{sup::stat_and_bounds}

\begin{figure*}[t]
\includegraphics[width=0.49\textwidth]{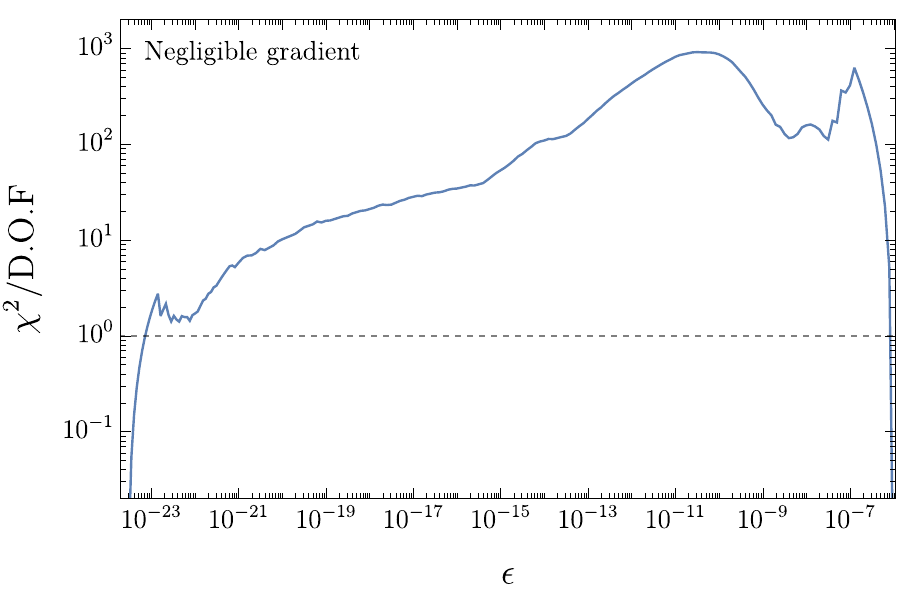}
\includegraphics[width=0.49\textwidth]{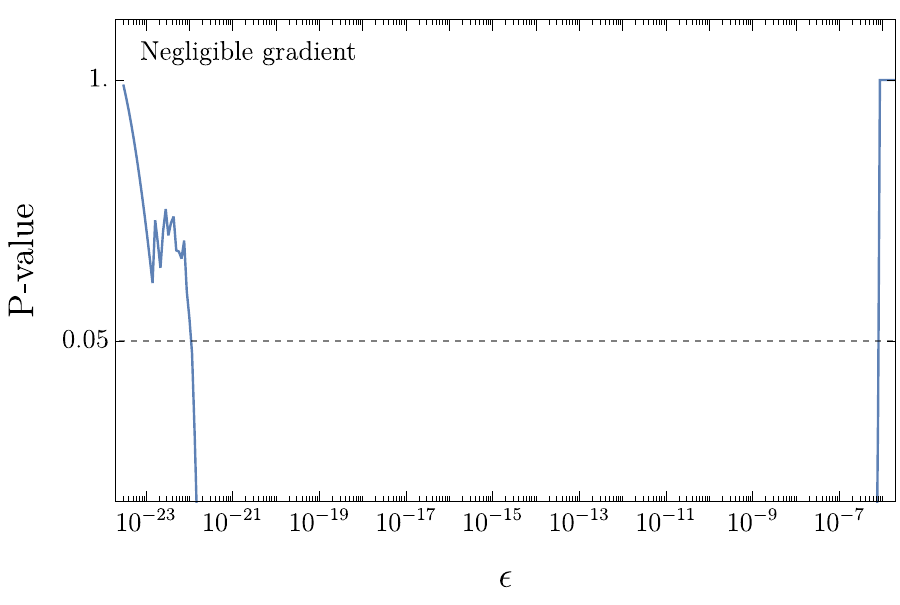}
\caption{Left panel: $\chi(\epsilon)$, as defined Eq.~(\ref{eq::chi_eps_def}) in the negligible gradient limit, normalized to the effective number of degrees of freedom $N-1 = 294$, as a function of $\epsilon$.  
Right panel: The p-value as a function of $\epsilon$. For reference, we plot the $2\sigma$ threshold, equivalent to $p=0.05$, as a gray dashed line.}
\label{fig:stat_f0}
\end{figure*}

\begin{figure*}[t]
\includegraphics[width=0.49\textwidth]{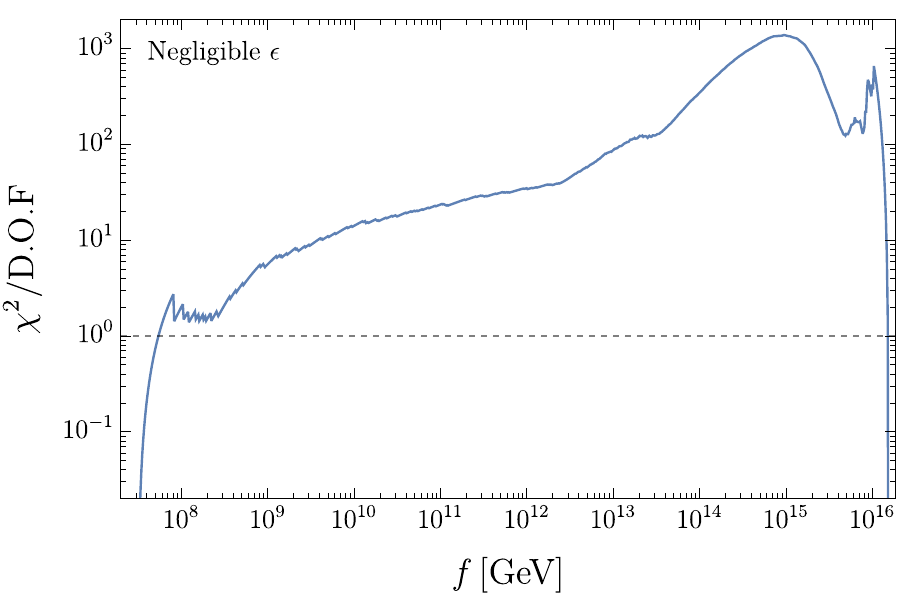}
\includegraphics[width=0.49\textwidth]{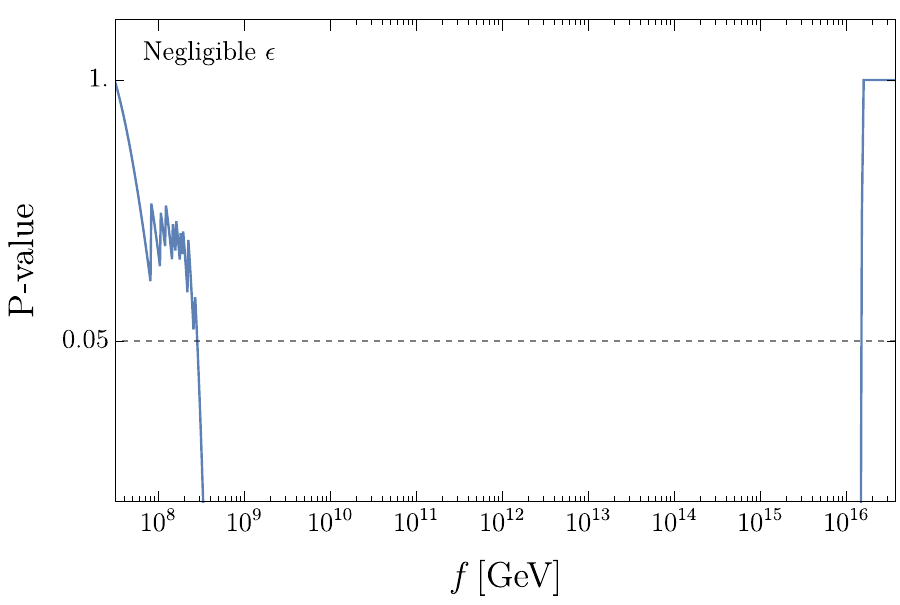}
\caption{Left panel: $\chi(f)$, as defined Eq.~(\ref{eq::chi_eps_def}) in the negligible $\epsilon$ limit, normalized to the effective number of degrees of freedom $N-1 = 294$, as a function of $f$.  
Right panel: The p-value as a function of $f$. For reference, we plot the $2\sigma$ threshold, equivalent to $p=0.05$, as a gray dashed line.}
\label{fig:stat_eps0}
\end{figure*}

\begin{figure*}[t]
  \centering
\includegraphics[width=0.4\textwidth]{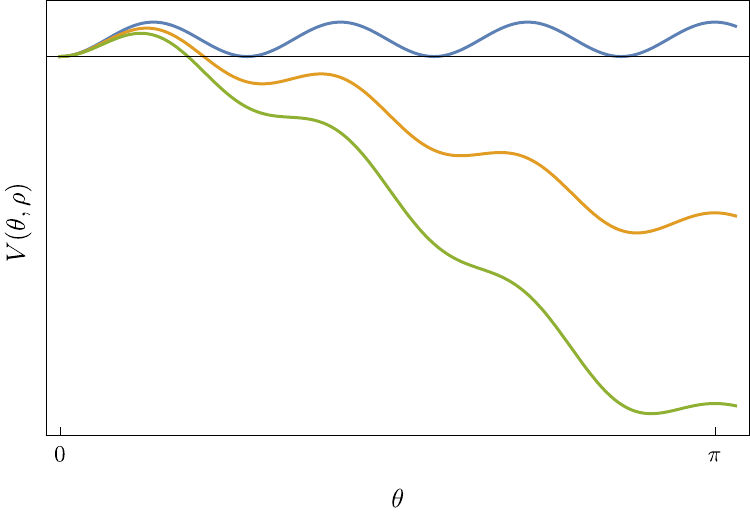}\hspace{+5pt}
\includegraphics[width=0.42\textwidth]{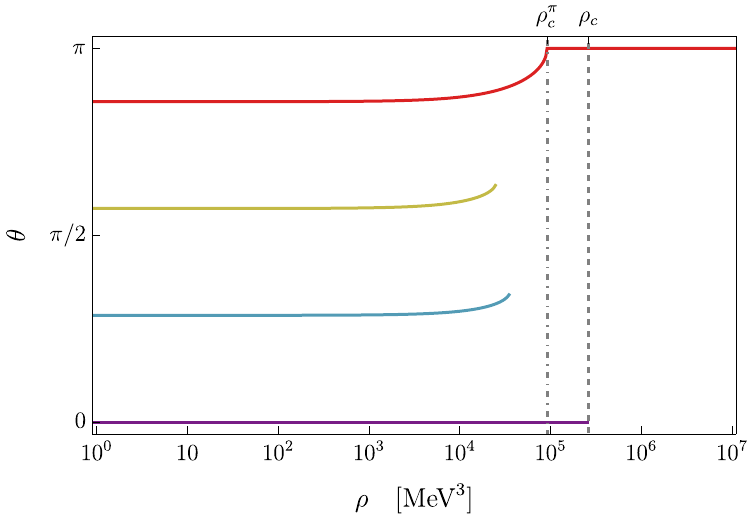}
       \caption{Left: Axion potential Eq.~\ref{eq:DenseZNpot} for $\mathcal{N}=7$, at zero density (blue), at finite density such that intermediate minima (at $\theta=2\pi/7,\theta=4\pi/7,\theta=6\pi/7$) are present (orange), and at a sub-critical finite density such that the intermediate minima are destabilized (green), see Eq.~\ref{eq:rhocritZN}. Right: $\theta(\rho)$ for $\mathcal{N}=7$.
    Rainbow color-coding marks the solutions to the $(7+1)/2=4$ minima from the one closest to $\pi$ (red) to the one at zero (violet).
    The dot-dashed and dashed gray lines correspond to $\rho_c^{\pi}$ and $\rho_c$ respectively.
    }
  \label{fig:DensPot}
\end{figure*}

The goal is to quantitatively determine how compatible are the observed WD radii with a gapped distribution.
Our working assumption is that the variance in the observed mass can be explained by varying other important properties of WDs, such as temperature and composition, which for simplicity we kept fixed. 
Therefore our focus is strictly on the radius distribution, and in order to determine the bounds on $\epsilon$ and $f$, we perform a 1D goodness-of-fit test on the radius axis.
Given the central values from a combined dataset of $N=295$ observed WD radii $\{r_i\}$ and their corresponding uncertainties~$\{ \sigma_i\}$\cite{B_dard_2017,tremblay_gaia_2016,2018MNRAS.479.1612J,Bond_2015,Bond_2017,10.1093/mnras/stx1522,Brown_2020}, we calculate the sum of squares, which we denote by $\chi$, for each point in the $\{ \epsilon,f\}$ plane
\begin{align}
\chi(\epsilon,f) = \sum_{i=1}^{N} \frac{D^2[(r_i),\Rmax(\epsilon,f),\Rmin(\epsilon,f)]}{ \sigma_i^2}\,,
\label{eq::chi_eps_def}
\end{align}

with the distance function
\begin{widetext}
\begin{equation}
D[(r_i),\Rmax(\epsilon,f),\Rmin(\epsilon,f)] = \begin{cases}
\text{min}\left[r_i-\Rmax(\epsilon,f),\Rmin(\epsilon,f)-r_i\right], \; &\;\;\;\;\;r_i \in [\Rmax(\epsilon,f),\Rmin(\epsilon,f)]
\\
0,\; & \;\;\;\;\text{otherwise}
\end{cases}\,.
\end{equation}
\end{widetext}

We use the numerically-calculated values for $\{\Rmax(\epsilon,f),\Rmin(\epsilon,f)\}$ where available, and otherwise use the verified analytical estimates as extrapolation, see right panel of Fig.~(\ref{fig:Rnum_Ranalytic_f0}) and Fig.~(\ref{fig:Rnum_Ranalytic_eps0}).

In Fig.~(\ref{fig:stat_f0}) and Fig.~(\ref{fig:stat_eps0}) we plot the results of our statistical analysis in the negligible gradient and negligible $\epsilon$ limits, respectively.  
In the left panels we plot $\chi$ normalized to the effective number of degrees of freedom $N-1$.

As a rough estimate, the range in which $\chi >1$ is considered incompatible with the gapped radii distribution hypothesis. 
A more refined statement can be made by calculating the corresponding p-values for each value of $\epsilon$ of $f$, shown in the right panel of Fig.~(\ref{fig:stat_f0}) and Fig.~(\ref{fig:stat_eps0}), respectively.  
We are able to exclude at the $2\sigma$ level the following interval in $\epsilon$
\begin{align}
2\times10^{-20} < \epsilon < 2\times 10^{-7} \;\;\;\; \text{(95\% CL)}\,.
\end{align}
We are able to exclude at the $2\sigma$ level the following interval in $f$,
\begin{align}
5.5\times10^{9} < f/\text{GeV} < 1.1\times 10^{16}\;\;\;\; \text{(95\% CL)}\,.
\end{align}

\begin{figure*}[t]
  \centering
  \includegraphics[width=0.45\textwidth]{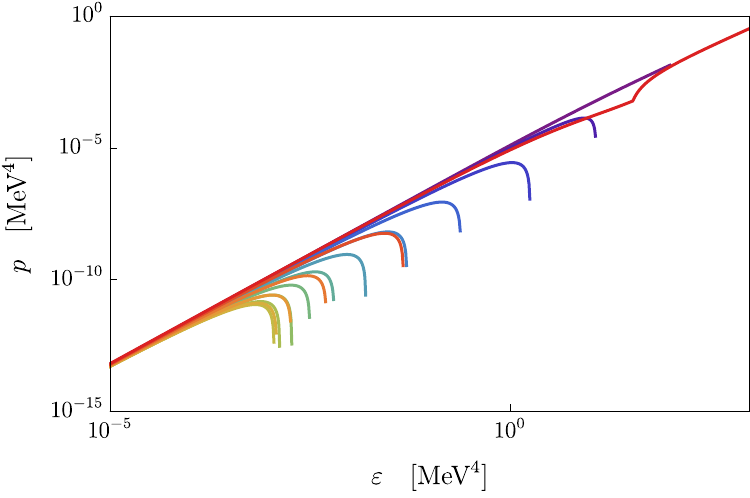}
\includegraphics[width=0.455\textwidth]{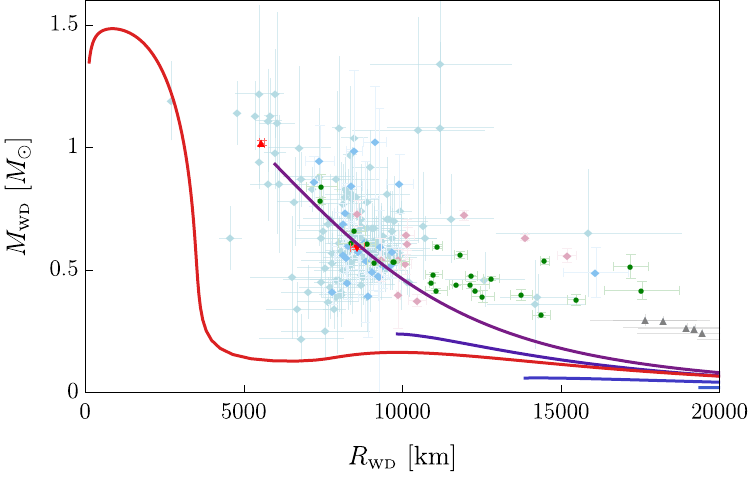}
\caption{Left: Pressure as function of the energy density for $\mathcal{N}=31$. Rainbow color-coding marks the solutions to the $(31+1)/2=17$ minima from the one closes to $\pi$ (most red) to the one at zero (most violet).
The red curve never experiences negative pressure.
While the pressure decreases due to the displacement of $\theta$ from $30\pi/31$, the Fermi pressure, due to electrons becoming relativistic, keeps the total pressure positive. 
Right: $M\,$-$\,R$ curves for $\mathcal{N}=31$ in the same color-coding.}
\label{fig:lowNMR2}
\end{figure*}    

\begin{figure*}[t]
  \centering
  \includegraphics[width=0.45\textwidth]
{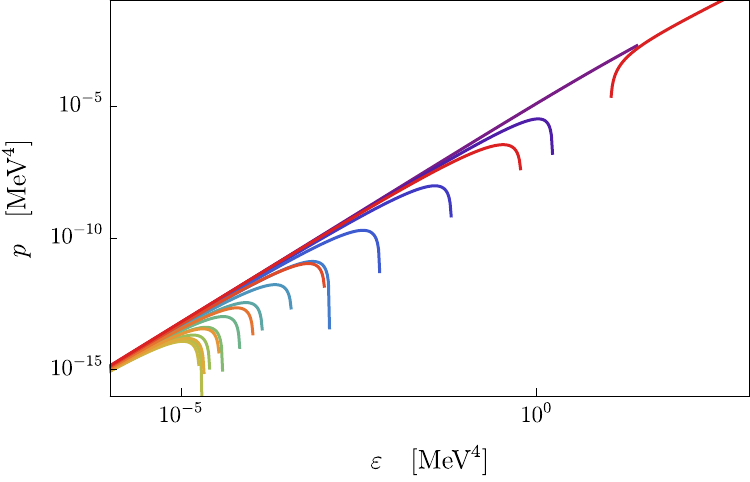}
\includegraphics[width=0.455\textwidth]{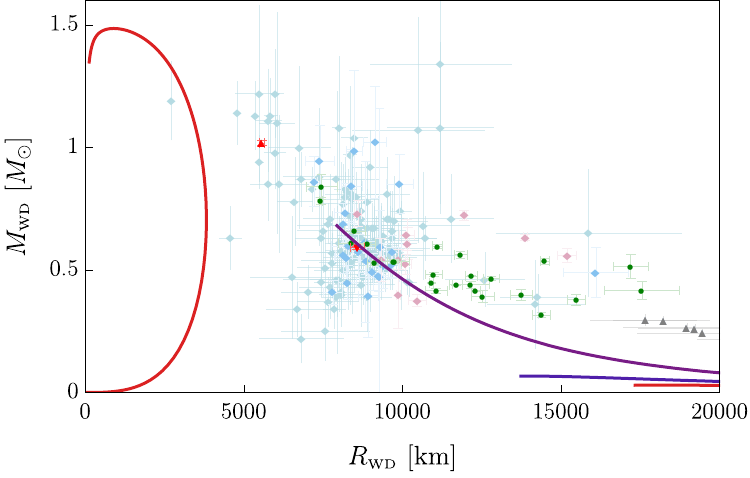}
    \caption{Left: Pressure as function of the energy density for $\mathcal{N}=33$. The rainbow color-coding marks the different branches of the EOS for the $(33+1)/2=17$ minima of $\theta$ with the ones closer to $\pi$ being more red. 
    Importantly, here the branch closest to $\pi$ experiences an instability. 
    Right: $M-R$ curves for $\mathcal{N}=33$ in blue (SBO) and orange (meta-stable) and for the corresponding $\epsilon$ (\Eq{eq:epsNrel}) in green (SBO) and red (meta-stable). }
  \label{fig:largeNMR1}
\end{figure*}

\section{$Z_\mathcal{N}$ axion}\label{app:ZN}

The potential of lighter than expected QCD axions due to a $Z_\mathcal{N}$ discrete symmetry reads, in the large $\mathcal{N}$ limit \cite{DiLuzio:2021pxd},
\begin{equation}
\label{eq:RingPot}
V_\mathcal{N}(\theta)=-m_{\pi}^2f_{\pi}^2\sqrt{\frac{1-z}{1+z}\frac{1}{\pi \mathcal{N}}}z^\mathcal{N}\left[\cos\left(\mathcal{N}\theta\right)-1\right],
\end{equation}
where 
$\mathcal{N}\gg 1$ is odd and $z=m_u/m_d$. The corresponding axion mass reads
\begin{equation}
m_a^2=\frac{\mathcal{N}^{3/2}z^\mathcal{N}\sqrt{1-z^2}}{\sqrt{\pi}(1+z)}\frac{m_{\pi}^2f_{\pi}^2}{f^2}.
\end{equation}
We can map the $Z_\mathcal{N}$ axion to the potential in \Eq{eq:potential} by requiring equal axion masses. 
We find the relation
\begin{equation}
\label{eq:epsNrel}
    \epsilon=\frac{1}{\sqrt{\pi}}\mathcal{N}^{3/2}z^{\mathcal{N}-1}(1+z)\sqrt{1-z^2}.
\end{equation}
While this relation fixes the axion mass to be the same in both models, the value of the potential at $\theta=\pi$ is different.
We find
\begin{equation}
    \label{eq:ZNpotPi}
    V_{\mathcal{N}}(\pi)\simeq\frac{1}{\mathcal{N}^2}V(\pi) \,,
\end{equation}
where $V(\theta)$ is the potential from \Eq{eq:potential}. 
Clearly, there is a large suppression for $\mathcal{N}\sim \mathcal{O}(10)$, which is the scenario we are interested in.

At finite density $\rho=k_F^3/3\pi^2$, the potential for the $Z_\mathcal{N}$ axion is given by \cite{DiLuzio:2021pxd}
\begin{equation}
\label{eq:DenseZNpot}
V_\mathcal{N}(\theta,\rho)=V_\mathcal{N}(\theta)+2\sigma_{\pi N}\rho\sqrt{1-\beta\sin^2\left(\frac{\theta}{2}\right)} \,,
\end{equation}
where $\beta=4z/(1+z)^2$.
The critical density, at which the minimum at the origin is destabilized, i.e.~$\partial^2_\theta V_\mathcal{N}(0,\rho_c)=0$ is
\begin{equation}
\label{eq:rhocritZN}
\rho_c=\frac{m_{\pi}^2f_{\pi}^2}{2\sqrt{\pi}\sigma_{\pi N}}\mathcal{N}^{3/2}z^{\mathcal{N}-1}(1+z)\sqrt{1-z^2} \,,
\end{equation}
which matches the critical density for the potential in Eq.~(\ref{eq:potential}), $\rho_c=\epsilon m_\pi^2f_\pi^2/2\sigma_{\pi N}$, with $\epsilon$ given by Eq.~(\ref{eq:epsNrel}).
At densities below the critical density, $V_\mathcal{N}(\theta,\rho)$ has $(\mathcal{N}+1)/2$ minima in the interval $\theta\in \left[0,\pi\right]$, which we label by $k=1,\dots,(\mathcal{N}+1)/2$.

At non-zero densities, minima with $k>1$ have lower energy than the minimum at $\theta=0$ (or $k=1$), but get destabilized at lower densities than the minimum at $\theta=0$.
In particular, we would like to point out that $\theta=\pi$ is initially not the lowest energy configuration since it remains a maximum for these intermediate densities, hence it is not preferred over e.g.~$\theta=(\mathcal{N}-1)\pi/\mathcal{N}$.
We solve $\partial^2_\theta V_\mathcal{N}(\pi,\rho_c^{\pi})=0$, and find that the density at which $\theta=\pi$ becomes the lowest energy configuration is
\begin{equation}
\label{eq:rhocritPi}
    \rho_c^{\pi}=\frac{1-z}{1+z}\rho_c
    \,.
\end{equation}
At densities $\rho_c^\pi\le\rho\le\rho_c$, two minima exist, at $\theta=0$ and $\theta=\pi$, while at densities $\rho>\rho_c$ the only remaining minimum is at $\theta=\pi$.

We show an $\mathcal{N}=7$ potential for zero and sub-critical densities in the left panel of Fig.~\ref{fig:DensPot}. 
In the right panel we show solutions $\theta(\rho)$ that sit in the $(\mathcal{N}+1)/2$ minima between $\theta=0$ and $\theta=\pi$ for $\mathcal{N}=7$.

In the negligible gradient limit, we can derive the EOS as described in the main text, i.e.~by solving \Eq{eq:minimum}.
At sub-critical densities we find 
$(\mathcal{N}+1)/2$ independent branches of the EOS, one for each meta-stable minimum.
The field value of $\theta$ increases with $k$ and so does the reduction of the mass.
The lowest energy configuration is therefore the branch that starts at $\theta=(\mathcal{N}-1)\pi/\mathcal{N}$ at 
low densities.

Solving the system for all branches of the EOS, we generally find a qualitatively distinct behaviour at small and large values of  $\mathcal{N}$.\\

\begin{figure*}[t]
  \centering
\includegraphics[width=0.45\textwidth]{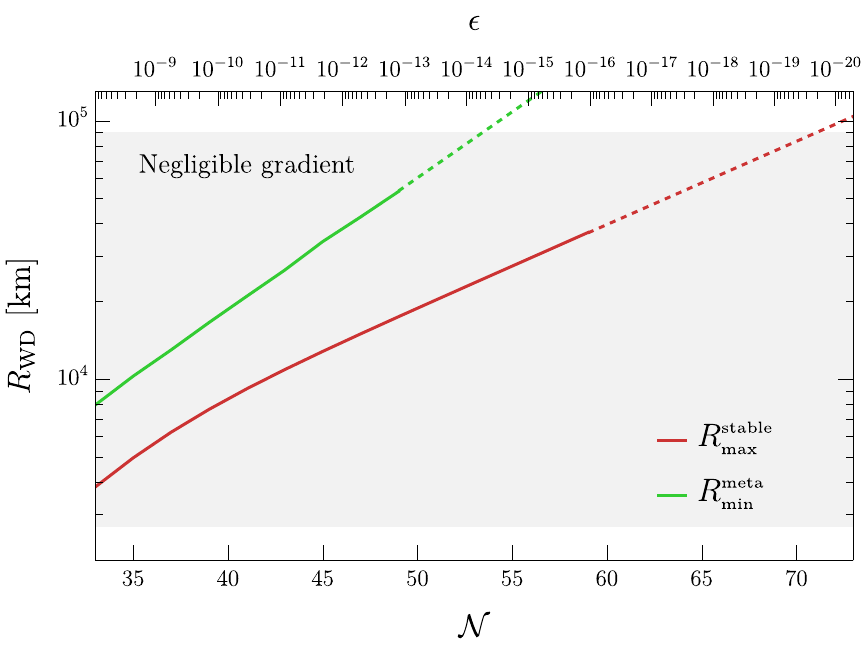}
\includegraphics[width=0.45\textwidth]{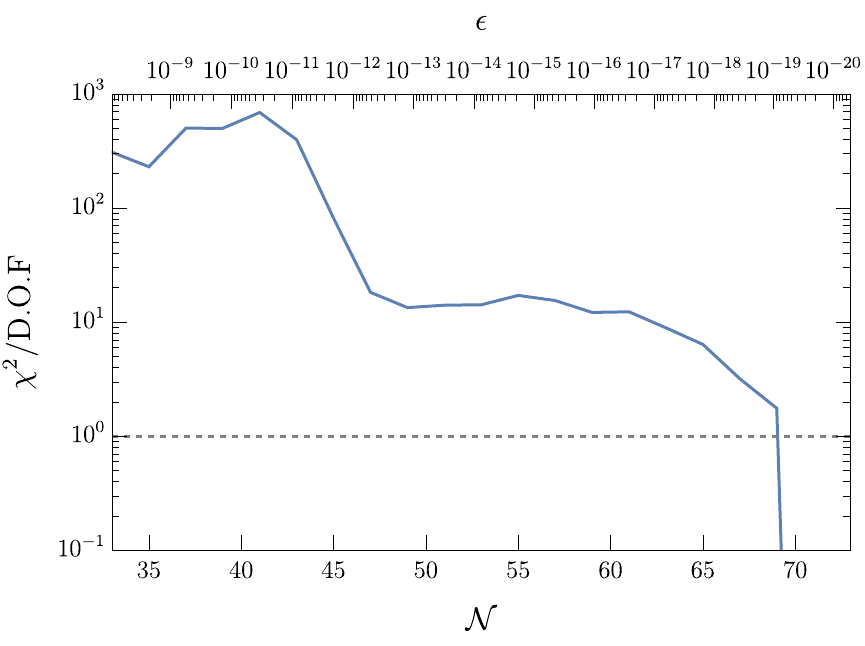}
     \caption{Left: Radius gap as a function of $\mathcal{N}$.
    Gray shaded area marks radii populated by data. Right: $\chi^2$ values over data points in the gap as a function of $\mathcal{N}$.}
  \label{fig:GapAndStat}
\end{figure*}

\subsection{Small $\mathcal{N}$}
At small $\mathcal{N}$, i.e. $\mathcal{N}\leq 31$, the electron Fermi pressure dominates over the negative contribution from the potential, such that the total pressure stays positive.
This is because electrons become relativistic at the relevant densities, since 
the lower 
the value of
$\mathcal{N}$, the larger the critical density, see \Eq{eq:rhocritZN}. 

Therefore, we do not find a gap in the EOS, which is qualitatively different from what we found for the potential in \Eq{eq:potential}.
We show the EOS, i.e.~$p(\varepsilon)$, in the left panel of \Fig{fig:lowNMR2} for $\mathcal{N}=31$.
Note that the EOS for the branch that sits closest to $\pi$ (red curve) does not experience a thermodynamic instability for any density, given that the pressure stays a monotonically increasing function of the energy density.
Nevertheless, the onset of the non-zero potential once the field starts to be displaced reduces the pressure significantly, leading to a much softer EOS at large densities than the EOS for $\theta=0$. 

In the zero gradient limit, the $M\,$-$\,R$
curve is readily found by solving the regular TOV equations \Eq{eq:TOV} with the prescribed EOS.
As expected from the EOS, we find continuous $M\,$-$\,R$ curves. 
For very low values of $\mathcal{N}$, sourcing happens only for the densest WDs.
We show the family of $M\,$-$\,R$ branches in the right panel of \Fig{fig:lowNMR2} for $\mathcal{N}=31$.

Even though the $(\mathcal{N}-1)\pi/\mathcal{N}$ (i.e.~$k=(\mathcal{N}+1)/2$) branch encounters an instability in the $M\,$-$\,R$ curve  at intermediate radii (where $\partial M_\text{\tiny WD}/\partial \rho_0<0$), the $\theta=0$ branch covers  these radii. 

Since we are agnostic about the formation mechanism of the star, and thus of the branch it ends up after formation, a $\chi^2$ in radii is not enough to exclude this prediction.
In principle an exclusion might be possible for the highest $\mathcal{N}$ in this regime, i.e.  $\mathcal{N} = 31$. This requires a more sophisticated statistical analysis than done for the gapped curve and is beyond the scope of this work.

\begin{figure*}[t]
\centering
\includegraphics[scale=0.5977]{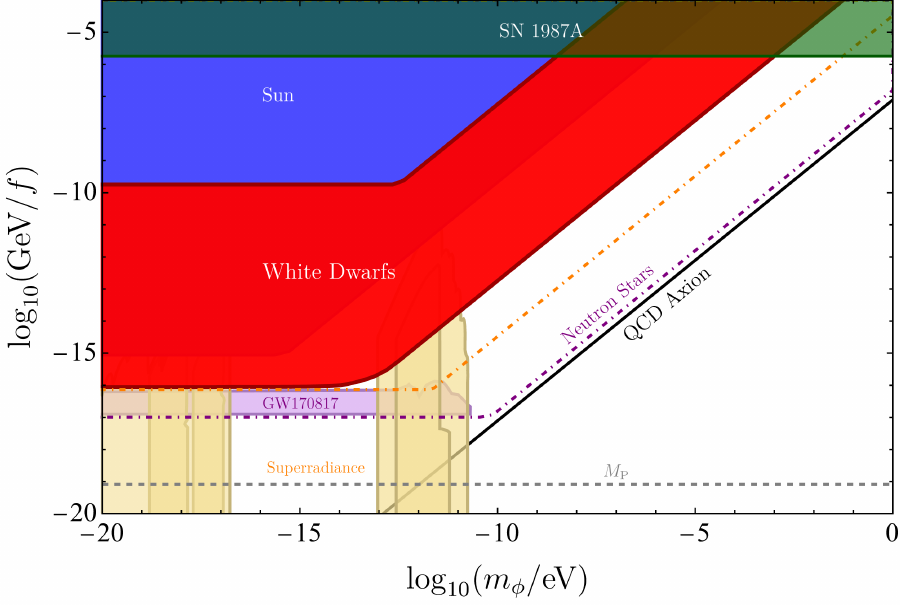}
\caption{\label{fig:MoneyplotZN} Constraints and future projections on the axion parameter space for the $Z_\mathcal{N}$-model. Same as in \Fig{fig:Moneyplot}.}
\end{figure*}

\subsection{Large $\mathcal{N}$}

At large $\mathcal{N}$ ,
i.e. $\mathcal{N}\geq 33$, we find that the negative contribution of the potential dominates over the electron Fermi pressure.
This implies a thermodynamic instability and a gap in the EOS, associated with a new ground state with density $\rho^* > \rho_c^\pi$. 
Contrary to the case of \Eq{eq:potential} (see \Fig{fig:EOSplot}), this gap can be covered by the $\theta = 0$ branch. 
This is because the $\theta=(\mathcal{N}-1)\pi/\mathcal{N}$ minimum disappears (and the one at $\theta=\pi$ appears) before the minimum at $\theta=0$ is destabilized, as follows from \Eq{eq:rhocritPi}. 
In \Fig{fig:largeNMR1}, for $\mathcal{N}=33$, it is shown that the region in which the instability occurs is covered for all densities by the meta-stable ($\theta=0$) branch. For such $\mathcal{N}$, the new ground state density $\rho^*$ lies below the critical density, as can be seen in the left panel of \Fig{fig:largeNMR1}.
For $\mathcal{N}\geqslant39$, we find instead $\rho^* > \rho_c$, a scenario that parallels that discussed in the main text.

Analytic estimates for the density of the new ground state can be found by solving
\begin{equation}
    p_e(\rho^*)-V_{\mathcal{N}}(\pi)=0 \,,
\end{equation}
in the non-relativistic and ultra-relativistic limit.
To leading order and neglecting $\mathcal{O}(1)$ numbers, these are given by
\begin{equation}
    \rho^*_\text{\tiny NR}\simeq
    \left(\frac{z^\mathcal{N}}{\sqrt{\mathcal{N}}}\right)^{3/5}
    \left(m_e m_\pi^2f_\pi^2\right)^{3/5} \,,\quad k_F\ll m_e \,,
\end{equation}
\begin{equation}
    \rho^*_\text{\tiny UR}\simeq
    \left(\frac{z^\mathcal{N}}{\sqrt{\mathcal{N}}}\right)^{3/4}
    \left(m_\pi^2f_\pi^2\right)^{3/4} \,,\quad k_F\gg m_e \,.
\end{equation}

As we have seen from the EOS, for larger values of $\mathcal{N}$ we find negative pressure phases in the $(\mathcal{N}-1)\pi/\mathcal{N}$ branch.
We therefore expect constant density self-bound objects (SBOs) and a gap in radii.
In the left panel of \Fig{fig:largeNMR1} we show the EOS for $\mathcal{N}=33$, which is the lowest $\mathcal{N}$ for which we find a new ground state. 
In the right panel of \Fig{fig:largeNMR1} we show the corresponding $M-R$ curves.
As can be seen, the new ground state leads to SBO solutions i.e.~the red curve that connects to $\MWD=0$ and $\RWD=0$.
This branch is disconnected by a gap in radii from the meta-stable branches at larger radii. 

In the left panel of \Fig{fig:GapAndStat}, we show the gap in radii as a function of $\mathcal{N}$ for $\mathcal{N}\geq33$ in .
In green we show the minimal radius of the meta-stable branch $R_\text{\tiny min}^{\text{\tiny meta}}$, where for very large radii we used the analytic estimate
\begin{equation}
    R_\text{\tiny min}^{\text{\tiny meta}}\simeq 10^4\left(\frac{2m_N\rho_c}{10^6\text{g}\,\text{cm}^{-3}}\right)^{-1/6}\text{km},
\end{equation}
see e.g. \cite{Shapiro:1983du}. 
In red we show the maximal radius of the stable branch $R_\text{\tiny max}^{\text{\tiny stable}}$ for which we use our numerical results and a similar analytic estimate
\begin{equation}
    R_\text{\tiny max}^{\text{\tiny stable}}\simeq \frac{\mpl{}}{m_N \sqrt{m_e}}\left(\frac{1}{\rho^*_\text{\tiny NR}}\right)^{1/6}.
\end{equation}
used.
In the right panel of \Fig{fig:GapAndStat}, we show the results of the $\chi^2$ analysis, described in \App{sup::stat_and_bounds}. 
As can be seen, at the $95\%$ confidence level, we are able to exclude $33\leq\mathcal{N}\leq 69$.

Last, we expect gradient effects to be of similar importance as in the detailed study above for only one minimum. 
Thus we expect also for the $Z_{\mathcal{N}}$ axion that the bound shuts down similarly around $f\sim 10^{16} \text{GeV}$. With an analogous statistical analysis as done above, we come to a slightly modified bound, as is shown in \Fig{fig:MoneyplotZN}.



\bibliography{WhiteDwarf}
\bibliographystyle{apsrev4-1}

\end{document}